%% file: main.tex
\documentclass{vgtc}                          
\ifpdf
  \pdfoutput=1\relax                   
  \usepackage{graphicx}                
  \DeclareGraphicsExtensions{.pdf,.png,.jpg,.jpeg} 
\else
  \usepackage{graphicx}                
  \DeclareGraphicsExtensions{.eps}     
\fi%

\graphicspath{{figures/}{pictures/}{images/}{./}} 

\usepackage{microtype}                 
\PassOptionsToPackage{warn}{textcomp}  
\usepackage{textcomp}                  
\usepackage{mathptmx}                  
\usepackage{times}                     
\usepackage{cite}                      
\usepackage{tabu}                      
\usepackage{booktabs}                  

\PassOptionsToPackage{hyphens}{url}\usepackage[hidelinks]{hyperref}
\usepackage{enumitem}
\usepackage{siunitx}
\usepackage[title]{appendix}
\usepackage{xspace}
\usepackage{amsmath, amsfonts}
\usepackage{comment}

\def\S{Sec.\xspace}
\def\A{Appendix\xspace}

\def\ie{\textit{i.e.,}\xspace}
\def\etal{\textit{et al.}\xspace}
\def\etc{\textit{etc.}\xspace}
\def\eg{\textit{e.g.,}\xspace}

\def\aka{\textit{a.k.a.}\xspace}
\def\vs{\textit{vs.}\xspace}
\def\first{\textit{first}\xspace}
\def\First{\textit{First}\xspace}
\def\Second{\textit{Second}\xspace}

\def\Finally{\textit{Finally}\xspace}

\def\Tool{\textit{Tool}\xspace}
\def\Hand{\textit{Hand}\xspace}
\def\Pen{\textit{Pen}\xspace}
\def\Controller{\textit{Controller}\xspace}
\def\Point{\textit{Point}\xspace}
\def\Pinch{\textit{Pinch}\xspace}
\def\MED{\textit{Medium}\xspace}
\def\SMA{\textit{Small}\xspace}
\def\LAG{\textit{Large}\xspace}
\def\LR{\textit{\mbox{Left--Right}}\xspace}
\def\FB{\textit{\mbox{Forward--Backward}}\xspace}
\def\UD{\textit{\mbox{Up--Down}}\xspace}
\newcommand{\N}{$12$}

\newcommand{\red}[1]{\textcolor{red}{#1}}

\newcommand{\presubsubsec}{\vspace{+.5em}}
\newcommand{\prepara}{\vspace{.20em}}

\newcommand{\postsec}{\vspace{0in}}
\newcommand{\presub}{\vspace{-0.1in}}
\newcommand{\postsub}{\vspace{-0.1in}}

\title{Investigating Input Modality and Task Geometry\\on Precision--first 3D Drawing in Virtual Reality}


\author{Chen Chen$^1$\thanks{corresponding author, e-mail: chenchen@ucsd.edu} \quad Matin Yarmand$^1$ \quad Zhuoqun Xu$^1$ \quad Varun Singh$^1$ \quad Yang Zhang$^2$ \quad Nadir Weibel$^1$\\ %
{\scriptsize \centering $^1$Computer Science and Engineering, University of California San Diego, La Jolla, CA, United States} \\ %
{\scriptsize \centering $^2$Electrical and Computer Engineering, University of California Los Angeles, Los Angeles, CA, United States}}

\input{00-abstract}

\CCScatlist{
  \CCScatTwelve{Human-centered computing}{Empirical studies in interaction design}{}{};
}

\begin{document}

\maketitle

\input{01-introduction}
\input{02-related}

\input{03-design}
\input{04-evaluations}
\input{05-discussion}

\input{06-limitation}
\input{07-conclusions}
\newpage
\begin{appendices}
\input{appendix}
\end{appendices}

\end{document}

%% file: 00-abstract.tex
\abstract{
Accurately drawing {\it non--planar} 3D curves in immersive Virtual Reality~(VR) is indispensable for many precise 3D tasks. 
However, due to lack of physical support, limited depth perception, and the non--planar nature of 3D curves, it is challenging to adjust mid--air strokes to achieve high precision. 
Instead of creating new interaction techniques, we investigated how task geometric shapes and input modalities affect precision--first drawing performance in a within--subject study~\mbox{(n = \N)} focusing on 3D target tracing in commercially available VR headsets.
We found that compared to using bare hands, VR controllers and pens yield nearly $30$\% of precision gain, and that the tasks with large curvature, forward--backward or left--right orientations perform best. 
We finally discuss opportunities for designing novel interaction techniques for precise 3D drawing. 
We believe that our work will benefit future research aiming to create usable toolboxes for precise 3D drawing.}

%% file: 01-introduction.tex
\section{introduction}\postsec\label{sec::intro}
Interactions in 3D immersive Virtual Environment~(VE) powered by today's Virtual Reality~(VR) have unlocked numerous applications in general work spaces~\cite{Chen2021} and healthcare~\cite{Chen2022, Chen2022VRcontour}.
The ability to directly draw in mid--air with physical tools and have traces appearing synchronously at the same physical location, has been identified as a key primitive to enable many drawing based applications in VR.
With such capabilities, for example, designers could efficiently externalize and communicate their ideas through 3D sketches~\cite{hutchins1985direct, Jacob1996}.
It is also important in numerous healthcare applications, where doctors need to precisely annotate on 3D medical structures~\cite{Chen2022, Chen2022VRcontour, Yarmand2022astro}.
%
However, it is challenging to enable precise drawing in unconstrained mid-air environments with commercially available input modality~(\eg~hands and VR controllers). 
For example, while designers are usually very excited at the prospect of using 3D sketching as a medium during the conceptualization phase, they constantly reported how they found it frustrating to draw in VE due to the lack of fine control over their strokes~\cite{Israel2009}. 

\begin{figure}[h]
    \centering
    \vspace{-0.10in}
    \includegraphics[width=0.5\textwidth]{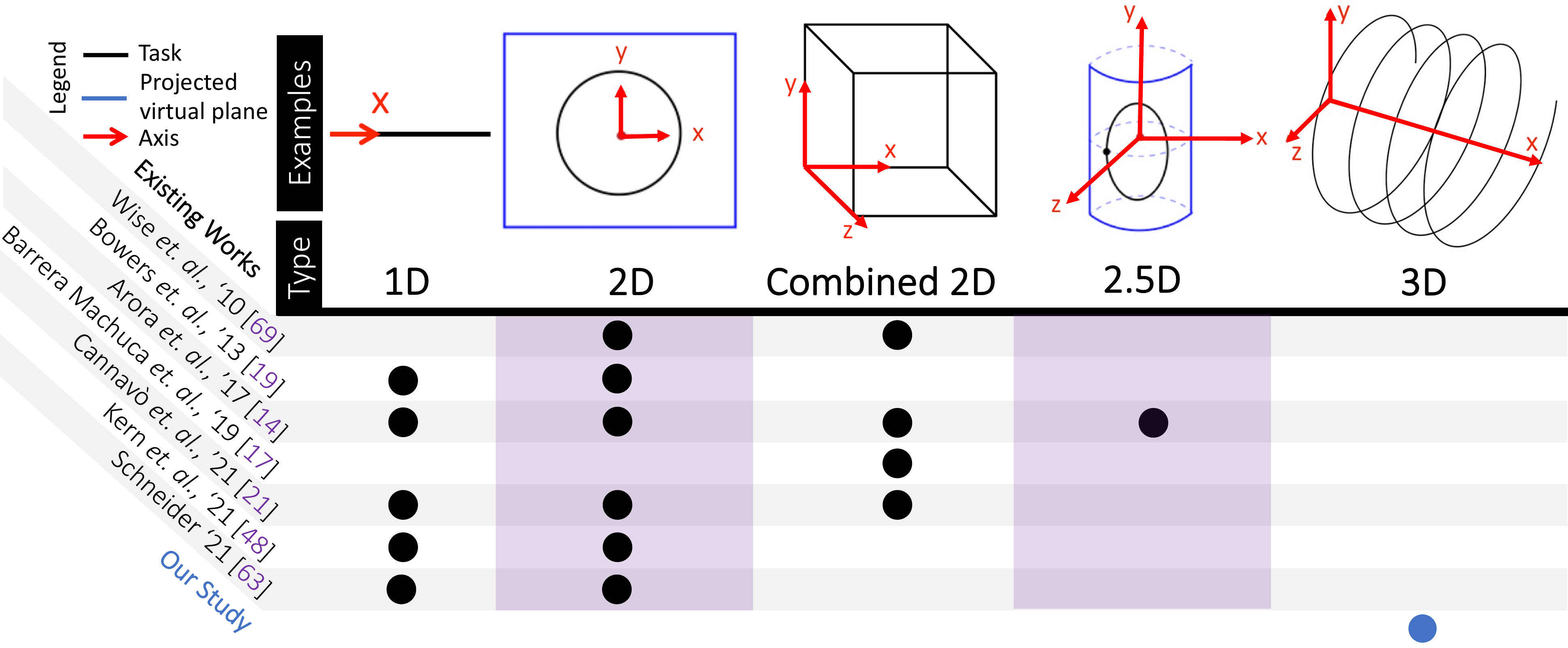}
    \vspace{-0.30in}
    \caption{Prior works that evaluates drawn tasks in VR. We focus on precision--first 3D drawing in VR. Demonstrative diagrams are revised from \cite{Arora2017}. Notably, ``combined 2D'' represents the curves that only consist of planar 2D curves and are not connected continuously, and ``2.5D'' represents the curves that are 2D curves projected on a 3D surface.}
    \label{fig::eval_related_ontology}
    \vspace{-0.15in}
\end{figure}

Although prior works have investigated the factors causing poor performance while sketching 2D curves on 2D surfaces~\cite{Helps2016, Accot1999, Meyer1994} and in 3D VE~\cite{Barrera2019, Arora2017, Cannavo2020}, understanding the performance of drawing {\it non--planar} 3D curves in immersive VE has not yet been explored~(see Fig.~\ref{fig::eval_related_ontology}). 
%
Unlike 2D curves, where all strokes are on the same virtual plane and can be described by a linear combination of two orthonormal basis, 3D curves pass through the  three-dimensional spaces and can \textit{only} be parameterized by three orthonormal basis.
This implies that while drawing on a particular surface, users need to adjust their strokes in the additional dimension, causing challenges of maintaining their drawing performance.
Therefore, it is important to understand how precise 3D drawing can be supported in VR, and how factors such as input modality and task geometric shapes could affect such delineation performance.

In this work, we \first investigate the performance of {\it precision--first 3D drawing}, referring to the interaction when users prioritize drawing precision over speed, while drawing 3D curves in VE.
Instead of creating new interaction techniques, we focus on understanding how the geometric shapes of 3D curves (\ie~curvatures, slopes, and orientations) and four widely used input methods (\ie~VR pen, VR controller, hand pointing, and pinching) can impact drawing performance, as measured by drawing error and error correction attempts (in the spatial domain), as well as speed (in the time domain).
%
%

We built proof-of-concept prototypes on \mbox{HTC Vive~\cite{HTCVivePro}} and \mbox{Oculus Quest~$2$~\cite{OculusQuest}}, to assess performance in a within--subject study \mbox{(n = \N)}. 
Our experiments focused on the analysis of $1,296$ trials and show that using a tool (\ie~the VR pen or controller) leads on average to $29.63$\% higher precision compared to hand gestures (\ie~pointing and pinching), and that using a VR Pen yields $13.11$\%  higher precision than using a controller. 
We also found that the large curvature, forward--backward or left--right orientation allow participants to generate the most precise drawing.
Finally, the qualitative evaluations of the user experience unveiled the importance of tools and occlusion during 3D precise drawing, and highlighted the demand of attention generated across the different geometric configurations.
In summary, our contribution is two fold: 
\begin{enumerate}[leftmargin=*, noitemsep, topsep=0pt, label=(\textbf{\arabic*})]

\item We characterize how \textbf{(a)} 3D geometric characteristics and \textbf{(b)} input methods, affect the
performance of drawing 3D curves in unconstrained 3D immersive environments (Sec.~\ref{sec::evaluation});

\item We discuss how our findings can be used for designing techniques to optimize drawing precision and experience in
unconstrained VE without introducing additional hardware (Sec.~\ref{sec::discussions});

\end{enumerate}

%% file: 02-related.tex
\section{Related Work}\postsec\label{sec::related}

\subsection{Input Methods for VR}\postsec\label{sec::related::tools}
%
Existing research has evaluated the usability of input devices for various VR tasks.
%
\mbox{Pham~\etal~\cite{Pham2019}} discovered that a pen--based interface is better than a controller for pointing tasks. 
%
\mbox{Batmaz~\etal~\cite{Batmaz2020}} demonstrated that using pen--like devices with precision grip could improve the performance for target selections.
\mbox{Li~\etal~\cite{Li2020Grip}} found that a tripod grip at the rear--end of the pen is the optimal posture for large ranges of motion.

Some prior research also investigated how to design novel stylus for VR (\eg~Flashpen~\cite{Romat2021} and OVR stylus~\cite{Jackson2020}).
Commercial VR styluses, \eg~Logictech VR Ink~\cite{LogitechVRInk} and Wacom VR Pen~\cite{WacomVRPen}, also show the potential of using pens for various VR input tasks.
While the ergonomic designs of these VR pens are usually considered as the vendor's slogan for advertising performance merits~ (\eg~{\it``sketch in the air, just like they would with a real pen''}~\cite{WacomVRPen} and {\it``offers control and precision''}~\cite{LogitechVRInk}), it is unclear in what contexts and how much affordance these styluses can actually offer, particularly when it comes to precise 3D drawing tasks.

With new advancements in vision and deep learning research, mainstream headsets like the Oculus Quest $2$~\cite{OculusQuest} have enabled hand--based gesture tracking, which can deliver a more natural and convenient way of interactions~\cite{OculusHandTracking, Han2020}.
Prior works have demonstrated the effectiveness of using bare-hand, either as a general input method for freehand sketching~\cite{OculusHandTrackingGestures}, as a surface drawing system that uses hand gestures to create, and manipulate the 3D sketches~\cite{Schkolne2001}, or as a bi--manual interaction modality geared to enhance the sketching precision by overcoming the lack of physical support~\cite{Jiang2021}. 
Despite these efforts, research outlining the differences between using bare--hand and standard input devices in the context of 3D curves creation is heavily under--explored.

We address the existing gap by comparing precise--first 3D drawing performance with four input methods: VR controller, VR pen, hand--pointing, and hand--pinching.
%
Unlike prior efforts using self--fabricated VR styluses~\cite{Pham2019, Arora2017, Li2020Grip}, we adopted Logitech VR Ink~\cite{LogitechVRInk}, a commercially available VR stylus, throughout the study.
%


\subsection{Evaluations of Drawing Performance in VR}\postsec\label{sec::related::eval}
%
%
Fig.~\ref{fig::eval_related_ontology} shows an inventory diagram, outlining the current research gap in evaluating drawing performance in VR.
%
%
For example, \mbox{Arora~\etal} evaluated sketching performance using pens and controllers, and found that the lack of physical surface support and the limited depth perceptions can cause sketching inaccuracies~\cite{Arora2017}.
%
%
Cannav{\`o}~\cite{Cannavo2020} investigated the performance of drawing 2D curves using the tasks designed by \cite{Arora2017}, but by using a consumer--grade Logitech VR Ink pen and the HTC Vive controller. 
Bowers~\etal~\cite{Bowers2021} compared line drawing performance with and without physical support on planar virtual canvases, and found that while the addition of a physical surface increased performance in terms of stroke speed and accuracy in some tasks, most participants still preferred to use air--drawing.
Kern~\etal~\cite{Kern2021} and Schneider~\etal~\cite{Schneider2021} focused on 2D interactions in 3D space with physical surface support.
%
%
While these works reported sketching performance when using a 2D surface in the 3D space, practical drawing in 3D often consists of 3D curves, which can only be modelled by a linear combinations of three orthonormal basis~(\ie~passing through the 3D space).
These are more challenging and cannot be supported by a planar (or curved planar) virtual surface.
%
Although Arora~\etal~\cite{Arora2017} included tasks with a few ``2.5D'' curves, \ie~a circle projected on a curved surface (see Fig.~\ref{fig::eval_related_ontology}), creating and interacting with those curves is fundamentally different than engaging with true 3D curves that could not be described by any planes (either curved or non--curved) and could only be modelled by three orthonormal bases.
Additionally, this work~\cite{Arora2017} did not focus on the impact of various 3D curve parameters~(\eg~curvature).

Few existing works focused on volumetric shapes composed of ``combined 2D'' curves, which include planar 2D curves that are not connected continuously (\eg~the skeleton of a cube). 
For instance, \mbox{Wise~\etal~\cite{Wiese2010}} focused on cognitive processes and found that sketching volumetric shapes consumes more time, and it is more error prone when compared to sketching flat shapes.
They also suggested how this is due to the additional interaction dimension where participants needed to make connections between several 2D sub--objects.
\mbox{Barrera Machuca~\etal~\cite{Barrera2019}} focused on the effects of spatial ability and indicated that participants' spatial ability affects the shapes of the line drawn, yet the line precision is less affected.
However, the ``combined 2D'' volumetric shape is different than actual 3D curves. Using  combined 2D curves implicitly reduces the task difficulty, leaving many research questions unanswered in terms of drawing in true 3D.
%

\begin{figure*}[t]
    \vspace{-0.15in}
    \centering
    \includegraphics[width=\textwidth]{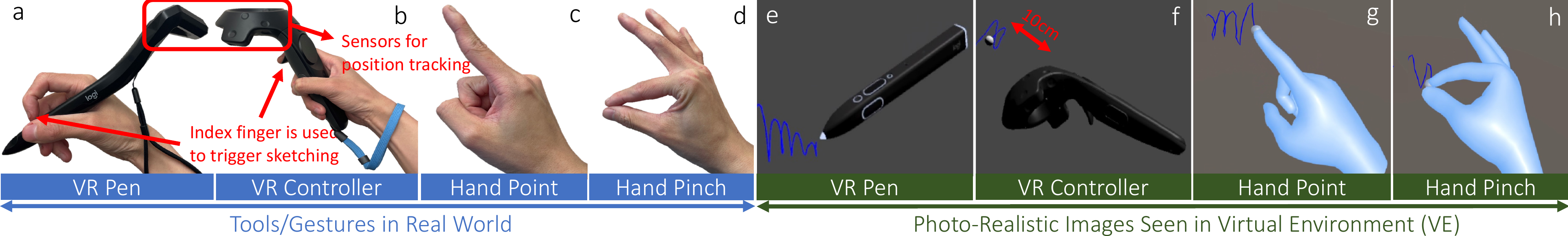}
    \vspace{-0.28in}
    \caption{We consider four input methods. On the left (a -- d) we show the tool-based and hand-based gestures in real world; on the right (e -- h) we show the photo-realistic prefabs that participants saw in the VR scene (the blue curves are drawn by the participants).}
    \vspace{-0.23in}
    \label{fig::input_methods}
\end{figure*}

\subsection{Techniques for Assisting Drawing in VR}\postsec\label{sec::related::sys}
Existing works also focused on assistive techniques to enhance 3D drawing performance. 
%
The first thread of systems attempted to use haptic feedback to enhance drawing performance.
%
%
For example, \mbox{Wacker~\etal~\cite{Wacker2018PhysicalGuides}} focused on the haptic feedback provided by actual physical interaction with a surface.
%
%
\mbox{Elsayed~\etal~\cite{Elsayed2020}} used pneumatic force and vibrotactile haptic to simulate the contact pressure of the stylus against sketching surface, and mimic the textures, respectively.
\mbox{Keefe~\etal~\cite{Keefe2007}} and \mbox{Mohanty~\etal~\cite{Mohanty2019}} looked into the effectiveness of incorporating a phantom haptic device.
%
%
%
Others (\eg~Arora~\etal~\cite{Arora2018}) integrated existing 2D sketching tools (\eg~tablet) into VR, to provide haptic feeling and physical support.
%
%
%
Although additional support using 2D surfaces can help with sketching 2D curves, it does not help 3D drawing and increases hardware cost and complexity, ultimately impacting the user experience.
We attempt to understand the performance of different {\it commercially available} and {\it widely used} input methods in 3D drawing tasks contexts.

Another thread is to design assistive visual elements to overcome the challenges introduced by the lack of depth perception.
Early works from \mbox{Grossman~\etal~\cite{Grossman2002, Grossman2001}} advocated the idea of leveraging virtual cross--sectional planes for drawing 2D curves while constructing 3D non--planar curves.
\mbox{Arora~\etal~\cite{Arora2017}} introduced visual feedback in terms of grids or scaffolding curves, and showed that this approach increases accuracy by $17$\% and $57$\% respectively. 
Smart3DGuides~\cite{Machuca2019} looked at the effectiveness of relying solely on visual guidance to increase the accuracy of the drawing shape likeness.
Multiplanes~\cite{Barrera2018Multiplanes} demonstrated a 3D drawing system that introduces virtual snapping planes and beautification triggers that are automatically generated based on previous and current strokes, as well as current controller poses.
Commercially available 3D modelling applications such as \href{https://www.gravitysketch.com}{Gravity Sketch}~\cite{GravitySketch} also used the snapping techniques, where the virtual snapping grids are used as the visual reference to help users better sketch the 2D curves.
While adding additional assistive visual components in the VR scene could help increase drawing performance, the immersive experience might be sacrificed~\cite{Bowman2007}.
In addition, the UI might become disorganized and overwhelming due to the additional visual components~\cite{Shneiderman2016}.

%% file: 03-design.tex
\begin{figure}[b]
\vspace{-0.2in}
    \includegraphics[width=0.5\textwidth]{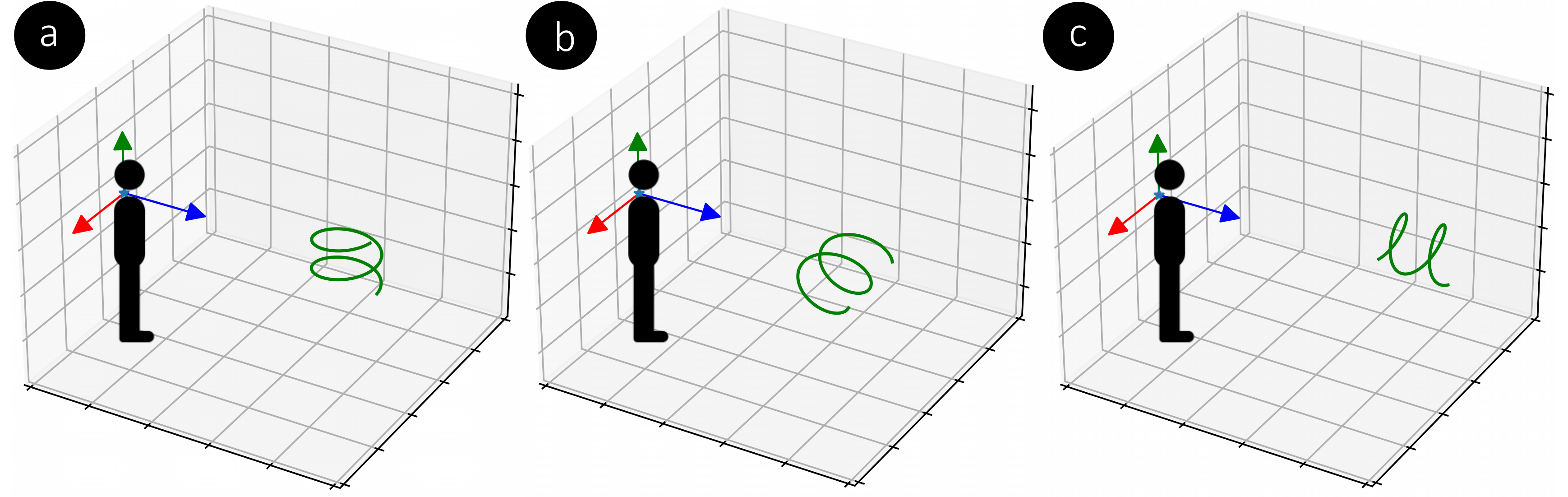}
    \vspace{-0.30in}
    \caption{The three types of orientations with respect to the participant: (a) \UD; (b) \LR; (c) \FB.}
    \vspace{-0.3in}
    \label{fig::task_property}
\end{figure}

\setlength{\abovedisplayskip}{0pt} 
\setlength{\abovedisplayshortskip}{0pt}

\section{Methods}\label{sec::design}
%

\subsection{Participants}\label{sec::method::participants}
We recruited \N~participants ($3$ females and $9$ males, age: \mbox{$\mu$ = $25$}, \mbox{$\sigma$ = $2.73$}). 
%
%
The average arm's length of recruited participants was $65.30cm$~($\sigma$ = $2.90cm$).
%
%
%
%
All participants were right--handed, and were instructed to complete the tasks with their dominant hand.
%
Our study was approved by the Institutional Review Boards~(IRB).

\subsection{Task Design}\label{sec::method::task_design}
%
We aim to measure performance when drawing continuous 3D curves in VE.  
We used helical curve as the primitive building block for designing our precision--first 3D drawing tasks for two reasons: 
%
%
%
\First, the helical curve mathematically represents one of the simplest continuous 3D curves with constant curvature, which can be easily described by only two parameters~(\ie~curvature and slope). 
This effectively helped us better design the controlled experiment.
\Second, the helical curve has been widely used in popular CAD tools~(\eg~Autodesk Inventor~\cite{AutoDesk3DHelix}) as the primitive building block for designers to create 3D prototypes.
With this rationale, we designed a within--participant study protocol with four distinct variables and a total of $108$~drawing combinations: orientation~(with three parameters), curvature~(with three parameters), slope~(with three parameters), and input method~(with four parameters).
%
%
The mathematical modelling and derivations of the helical curves are described in \A~\ref{sec::app::helix}.

\prepara
\noindent\textbf{Orientation} --
The drawing tasks followed three types of orientations with respect to participants' shoulders: \UD, \LR and \FB~(see Fig.~\ref{fig::task_property}).
We positioned the center of each helical curve on the invisible hemisphere, centered around the shoulder~(the positioning of participants shoulder is determined during the calibration phase, as described in \S\ref{sec::method::procedure}). 
The \LR and \FB configurations can be generated by rotating the helical curve \ang{90} with respect to the {\it y} and {\it x} axes respectively.

\prepara
\noindent\textbf{Curvature and Slope} --
A 3D helical curve can be parameterized by its curvature and slope.
We devised three cases~(\ie~\SMA, \MED, and \LAG) for each of the variables, contributing to nine conditions: $4m^{-1}$, $8m^{-1}$, and $12m^{-1}$ for curvature, as well as $0.05$, $0.25$, and $0.45$ for slope.
The parameters were determined by ensuring that each target stroke is within all participants' reach, without having to physically move from the standing position.

\prepara
\noindent\textbf{Input Method --}
We base our evaluation on four input methods, including two 3D sketching tools (\ie~VR \Controller and VR \Pen) and two air--drawing hand gestures (\ie \Point and \Pinch). 
Fig.~\ref{fig::input_methods} demonstrates these input methods, along with the photo--realistic prefab shown in the VR scene.

\prepara
\noindent$\bullet$ {\bf Tool Based Inputs --}
We selected two commercially available VR input devices, the Logitech VR Ink~\cite{LogitechVRInk}~(hereafter referred as \Pen) and the HTC Vive Controller~\cite{ViveController} (hereafter referred as \Controller).
Participants were instructed to establish grip with the stylus using a \emph{tripod front-end gesture} (see Fig.~\ref{fig::input_methods}a).
According to~\cite{Li2020Grip}, the tripod front--end and rear--end gripping postures are the easiest for distant target selections, but the tripod rear--end would be impractical due to the bulky size of the tracking rack. 
We demonstrated the gesture by showing how to use thumb and middle finger to stabilize and control the stylus, and use the index finger to trigger drawing by pressing the primary button. 
All participants were comfortable with the grip before starting to work on their tasks.
%
Similar to the pen, we instructed participants to draw using the controller by pulling the trigger button using their index finger and \mbox{hold $+$ stabilize} the controller using the palm and rest of fingers~(see Fig.~\ref{fig::input_methods}b).

\prepara
\noindent$\bullet$ {\bf Hand Based Inputs --}
We included two types of air--drawing hand gestures: pointing and pinching.
The drawing would be triggered by a foot switch, aiming to prevent potential confounding factors associated with two hands interactions.
%
Typically the pointing gesture is preferred for fine grained direct manipulations~\cite{Aigner2012}.
To ensure consistency, we asked all participants to use their index finger for the hand pointing gesture. 
For fine--grained control over the drawing stroke, we encouraged the participants to incorporate different levels of finger dexterity~(\eg~bent or stretched index finger).
%
We also included the pinch gesture due to its popularity in mainstream AR/VR systems~\cite{OculusHandTracking, HoloLensPinch}.
We instructed participants to use the thumb--index pinch (Fig.~\ref{fig::input_methods}d).
While the pinch gesture offers less flexibility for finger dexterity, it provides a \emph{self-haptic}, tactile feedback during the interaction, which has been used in existing AR/VR systems~\cite{OculusHandTracking, Young2020}.

\subsection{Measures}\postsec\label{sec::design::measures}
We evaluate drawing performance using measures in spatial (MOD \& NLM) and temporal domains (MSP), in short:
%
%
%
%

\prepara
\noindent{\bf \underline{M}ean \underline{O}verall \underline{D}eviation~(MOD) --} 
We use MOD to measure the average error between strokes and target curves~\cite{Arora2017}. 
After removing all extreme outliers using an interquartile range~(IQR) criterion~\cite{Dekking2005}, we compute the mean of stroke--target distance (\ie~the Euclidean distance of each participant-drawn stroke to the target traces) (see equation~\ref{eqn::mod}, where stroke and target traces are denoted by $\boldsymbol{P}$ and $\boldsymbol{T}$, respectively; we used $d(p_i, T)$ to denote the $L2$ distance between stroke $p_i$ to target trace $\boldsymbol{T}$).


\vspace{-0.15in}
\begin{equation}
    MOD = \frac{1}{N} \sum_{i = 0}^{N - 1} [d (\overrightarrow{p_i},  \boldsymbol{T})] 
    \label{eqn::mod}
\end{equation}
\vspace{-0.15in}

\prepara
\noindent{\bf \underline{N}umber of \underline{L}ocal \underline{M}axima of Stroke Deviations~(NLM) --}
%
%
It is impractical for participants to position every strokes precisely on the target trace.
%
While drawing, participants might try to correct the errors by re--adjusting the stroke on top of the target trace.
We computed NLM as the metric to evaluate how frequently the participants attempted to correct the instantaneous drawing errors~\cite{Itaguchi2018} (see equation~\ref{eqn::nlm}, where $\mathbb{I} (\cdot)$ denotes the indicator function).
A higher NLM theoretically indicates that the stroke is difficult to be stabilized.

\vspace{-0.12in}
\begin{equation}
    NLM = \sum_{i = 1}^{N - 2} \mathbb{I}[d (\overrightarrow{p_i},  \boldsymbol{T}) > d (\overrightarrow{p_{i - 1}},  \boldsymbol{T})] \mathbb{I}[d (\overrightarrow{p_i},  \boldsymbol{T}) > d (\overrightarrow{p_{i + 1}},  \boldsymbol{T})] 
    \label{eqn::nlm}
\end{equation}
\vspace{-0.15in}

\prepara
\noindent{\bf \underline{M}ean Drawing \underline{SP}eed (MSP) --}
%
%
We calculated the instantaneous speed at every stroke, dividing the Euclidean distance~(from the previous point) by the elapsed time between two samples.  

\begin{figure}[b]
    \centering
    \vspace{-0.2in}
    \includegraphics[width=0.5\textwidth]{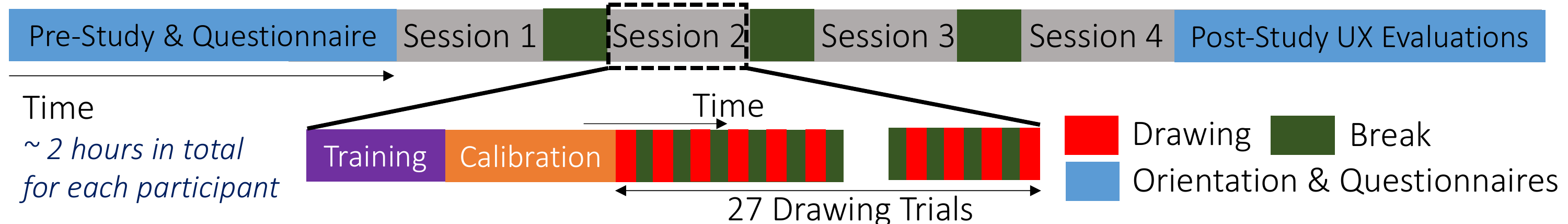}
    \vspace{-0.25in}
    \caption{Study timeline. During long breaks~($10 \sim 15$ min), participants were asked to take off the headset. During short breaks, participants were asked to rest their arms for $2 \sim 3$ seconds.}
    \vspace{-0.30in}
    \label{fig::study_timeline}
\end{figure}

\begin{figure*}[t]
    \centering
    \includegraphics[width=\textwidth]{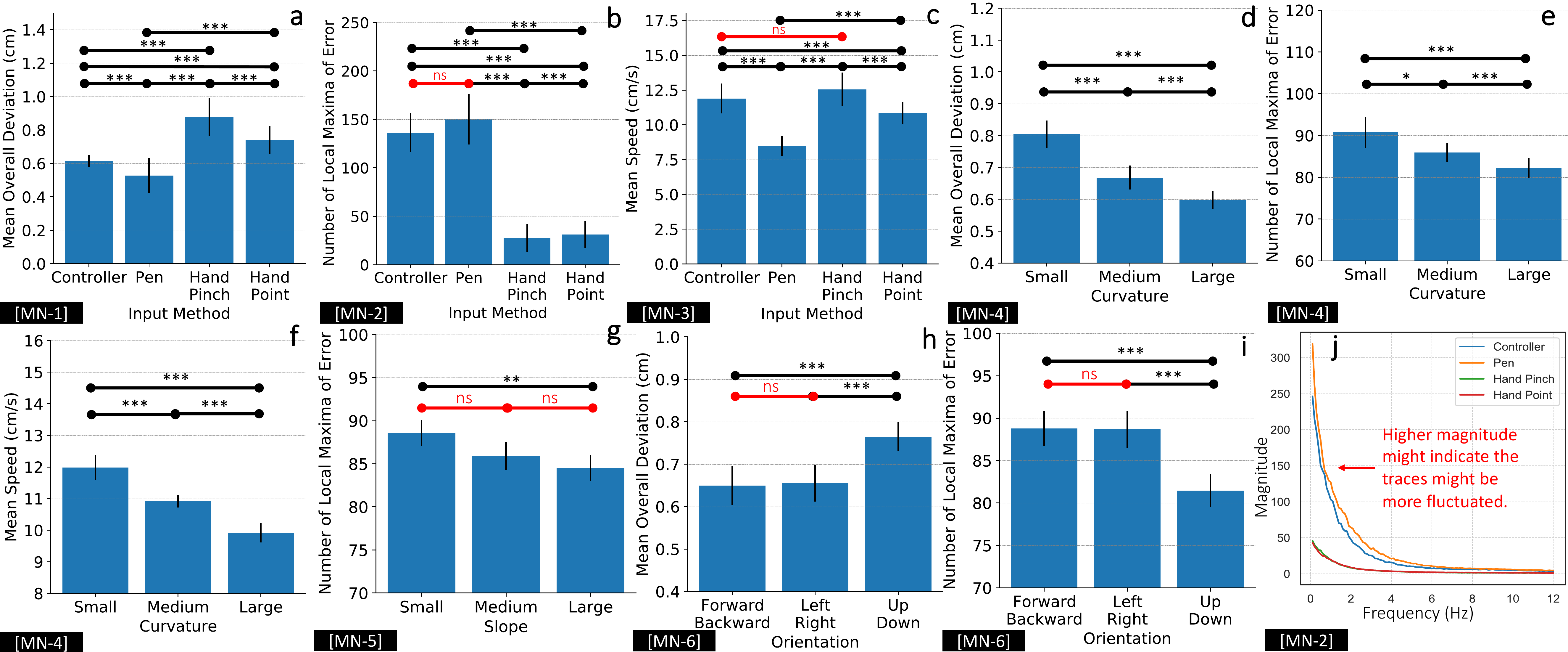}
    \vspace{-0.30in}
    \caption{(a -- i) Analysis of main effects~($* = p < .05$, $** = p < .01$, $*** = p < .001$), while the red colored ``$ns$'' indicates no statistical significance. (j) FFT of instantaneous stroke--target distance~(\aka~error). We only show the frequency components below $12$~Hz, as the frequency for one's tremors is usually within $8$ -- $12$Hz, while higher frequency components could be caused by positional tracking and is negligible in terms of human performance evaluations~\cite{Charles1999}. Indices of key takeaways are labelled at bottom left of each subplots.}
    \vspace{-0.23in}
    \label{fig::eval_main_effect}
\end{figure*}

\subsection{System and Implementation}\postsec\label{sec::methods::impl}
%

We implemented the testbed for controller and pen input using HTC Vive~\cite{HTCVivePro} and hand based input using Oculus Quest~$2$~\cite{OculusQuest}.
We used Vive for implementing the testbed of tool based input, because at the time of study the Oculus Quest~$2$ did not support an SDK for the Logitech VR stylus that we wanted to use to evaluate drawing precision. 
%
%
%
Additionally, using Vive for evaluating precision--first bare hand drawing was impractical due to the low accuracy of hand tracking. 
%
%
We therefore explored multiple tracking options and decided against using optical motion capture~(MoCap) system (\eg~OptiTrack) for three reasons.
\First, MoCap systems usually requires wearing smart gloves with Infrared~(IR) markers, which introduces additional weight and haptic feelings, and could cause non--negligible differences when compared to bare hand drawing.
\Second, the tracking performance of certain hand poses might degrade due to the occlusion of IR markers.
%
%
%
\Finally, we wanted to focus on participants' 3D drawing performance in readily accessible commercially available VR systems. 
The high costs, complicated setups, potential issues caused by reflections introduced by MoCap hinders the practical uses.
With considerations of commercial device availability, we eventually decided to prototype the hand-based testbed on Quest $2$ using the Oculus Integration SDK.

\subsection{Procedure}\postsec\label{sec::method::procedure}
%
%
%

Our experimental procedure consist of three major steps (Fig.~\ref{fig::study_timeline}):

\prepara
\noindent\textbf{Pre--Study and Questionnaire -- }
Participants were first asked to complete a pre--study questionnaire aiming to collect background information, and a $6$--min Paper Folding Test~(VZ-2)~\cite{paperfolding} to evaluate their spatial ability. 
%
All participants exhibited above-average spatial abilities, with an average test score of $92.92\%$ ($\sigma$ = $10.54\%$). 
%

\prepara\noindent\textbf{Evaluations -- }
We structured our evaluation around four sessions (see Fig.~\ref{fig::study_timeline}), with each pertaining to one input methods, chosen based on balanced Latin Square algorithm~\cite{Dekking2005}.
We chose to group trials by input method to minimize the time and effort for switching tools. 
Participants completed $27$ trials in each session with the order being randomized.
Each session consists of three phases: 

\prepara
\noindent$\bullet$\textbf{Phase 1: Training -- }
Each session was preceded with a training phase.
\First, the experimenter demonstrated the proper use of the equipment and the drawing tasks.
\Second, the experimenter helped participants to wear the headset and to try and draw until they felt comfortable with the environment and input method. On average, this process lasted around $5 \sim 10$~minutes.

\prepara
\noindent$\bullet$\textbf{Phase 2: Calibration -- }
%
Due to the diversity in participants' heights and arms' lengths, we incorporated a calibration phase before the main drawing tasks.
We used the calibration process to adjust the positions and orientations of the drawing tasks in the subsequent trials so that they appear at relatively constant locations across all participants.
Participants were asked to hold the controller, stand at a fixed location, and keep their arm straight and raised it to the forward, up, and right position, respectively.
%
%
While keeping the gesture constant for $5$~seconds, the system sampled and computed the mean location of the controller.
We finally used a hemisphere centered around participants' shoulder with a radius equal to the arm length to denote the reachable region by participants.
The calibration results were then used to generate subsequent trials.

\prepara
\noindent$\bullet$\textbf{Phase 3: Drawing -- }
Participants were finally asked to trace over the helical shapes that appeared in the VE using the associated input method.
A progress information panel was visible to help keep tracking the current state of the trial and also improve engagement~\cite{progressbarUsability}.
As we aim to evaluate the precision of the drawing performance, we asked participants to prioritize accuracy over speed (\ie~increase their speed \underline{\textit{only after}} participants believed they reached the highest possible accuracy).
Additionally, to avoid motion sickness and fatigue, the study session contained multiple short breaks~(between trials) and long breaks~(between sessions) (see  Fig.~\ref{fig::study_timeline}).
During the short breaks, participants put their arms down momentarily to relieve any fatigue. 
To avoid position inconsistency, we asked participants not to move away from the standing position.
After every session, participants took a longer break during which the headset and input devices~(if any) were removed.
On average, short breaks took approximately $2 \sim 3$  seconds and long break $10 \sim 15$ minutes.

\prepara
\noindent\textbf{Post-Study User Experience~(UX) Evaluations -- }
We finally evaluated participants' experience by asking them to fill out an experience evaluation questionnaire and think--aloud while recording their responses.
%
%
Our evaluation questions were revised from~\cite{Elsayed2020}.
As we focused on evaluating participants' experience with respect to a particular input techniques instead of the entire system experience, we excluded the questions of {\it system engagement} and {\it system convenience}.
We also added an additional assessment that focused on fatigue.
On a $5$--point Likert scale, participants indicated to what extent they agreed or disagreed with the following two statements:

\noindent$\bullet$ {\bf Confidence}: I am confident that I could draw the shape correctly.
\noindent$\bullet$ {\bf Fatigue}: I do \underline{\textit{not}} feel increased fatigue using this device.

%% file: 04-evaluations.tex
\begin{figure*}[t]
    \centering
    \includegraphics[width=\textwidth]{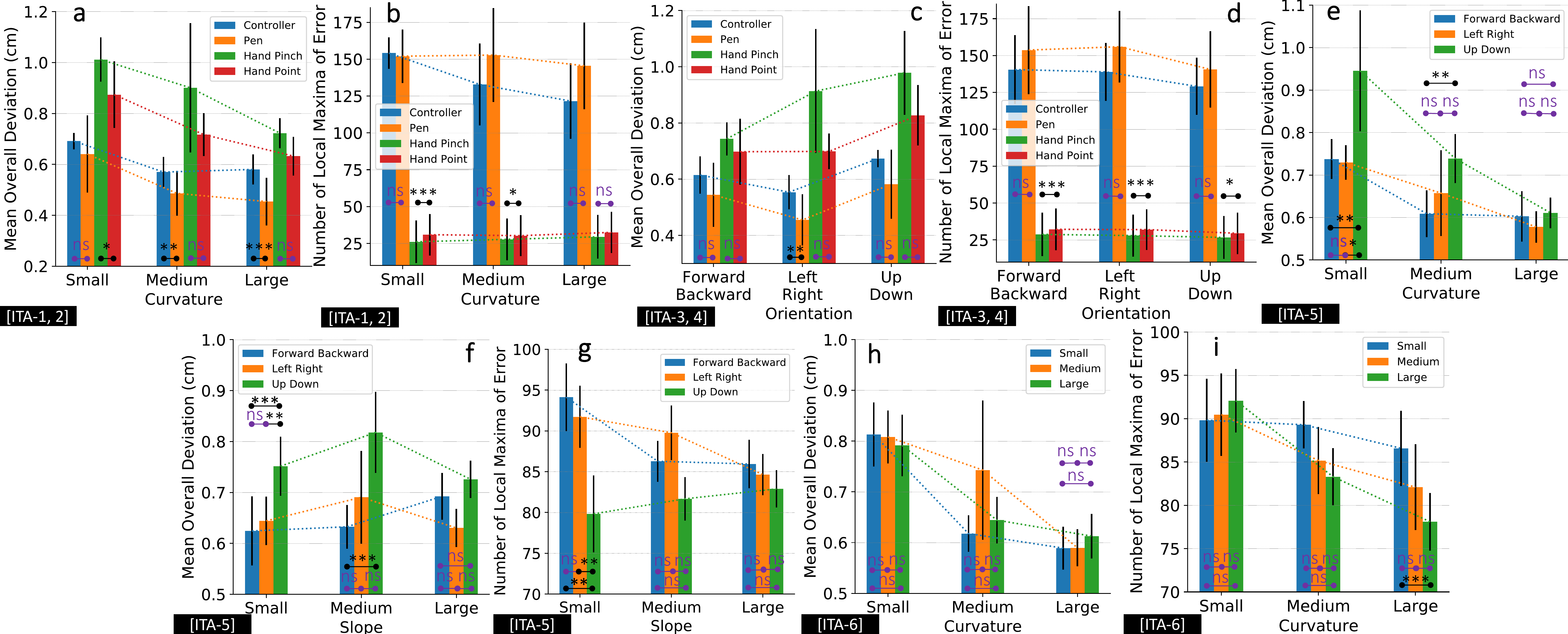}
    \vspace{-0.30in}
    \caption{Analysis of two--way interaction effects ($* = p < .05$, $** = p < .01$, $*** = p < .001$). The purple colored ``$ns$'' indicates no statistical significance. To enhance readability, we only annotate the contrast tests within each group (\ie~when the independent variables of the horizontal axis are identical). While visualizing the interaction effects involving input methods, we only annotate the contrast test results by comparing within \Tool or \Hand. Indices of key takeaways are labelled at bottom left of each subplots.}
    \vspace{-0.23in}
    \label{fig::interaction_effects}
\end{figure*}

\section{Results}\postsec\label{sec::evaluation}
%

This section outlines the key results, with \textbf{\mbox{[MN-\#]}} and \textbf{\mbox{[ITA-\#]}} referring \underline{m}ai\underline{n} and \underline{i}n\underline{t}er\underline{a}ction effects, \textbf{\mbox{[ITM-\#]}} referring \underline{i}n\underline{t}er\underline{m}ediate variables, and \textbf{\mbox{[SUB-\#]}} referring participants' \underline{sub}jective comments.

\subsection{Independent Variables}\postsec\label{sec::results::independent}
To examine the statistical significance of the effects of the four independent variables over the three measures, we first perform the normality check, and found that $39$, $9$, and $17$~(out of $108$) observations for MOD, NLM and MSP did not comply to the normal distribution~($p < .05$).
Therefore, we decided to process all data using Aligned Rank Transform~(ART)~\cite{art2011}, followed by a Repeated--Measure~(RM) Analysis of Variance~(ANOVA)~($\alpha = .05$).
The ART procedures for multifactor contrast test was used for post--hoc test, with Bonferroni corrections~\cite{artc2021}.
%
%
%
%
As for confidence intervals, we first transformed the data using Cousineau's method~\cite{Cousineau2005} to minimize the impacts of between--subject variations, and then computed the length of error bars using $95\%$ confidence intervals.

\presubsubsec
\noindent\underline{\bf Analysis of Main Effects}~\\
Main effect refers to the statistical relationship between the four input methods and the three measures, averaging across the level of the other independent variables.
Fig.~\ref{fig::eval_main_effect} visualizes the main effects that are considered statistical significant~($p < .05$).

\begin{figure*}[t]
    \centering
    \includegraphics[width=0.85\textwidth]{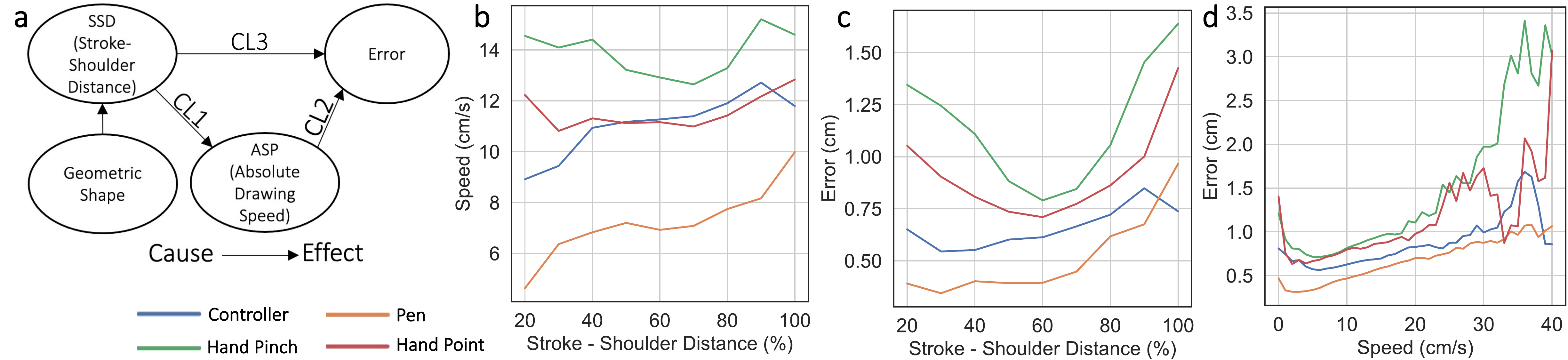}
    \vspace{-0.15in}
    \caption{(a) Causal model of the three variables of interests. (b -- d) Correlations between (b) ASP \vs~SSD, (c) Error \vs~SSD, and (d) Error \vs~ASP. For stroke samples with the same independent variable, we only visualize the average values of dependent variables.}
    \vspace{-0.23in}
    \label{fig::eval_intermediate}
\end{figure*}

\prepara
\noindent{\bf Input Method --}
%
RM--ANOVA indicates a strong statistical significance of input methods over MOD~($F_{3, 33} = 142.90$, $p < .001$, $\eta^2_p = 0.267$), NLM~($F_{3, 33} = 2299.12$, $p < .001$, $\eta^2_p = 0.854$) and MSP ($F_{3, 33} = 185.24$, $p < .001$, $\eta^2_p = 0.321$).
By examining \mbox{Fig.~\ref{fig::eval_main_effect}(a -- c)}, we conclude three key findings:

\prepara
\noindent $\bullet$ \emph{\mbox{\textbf{[MN-1]}} {\bf MOD --}}~ Using a \Tool (\Pen and \Controller) leads to $29.63$\% less MOD when compared to \Hand input ($\mu_{tool} = 5.7mm$ \vs $\mu_{hand} = 8.1mm$).
Considering just \Tool inputs, the MOD generated by the \Pen is $13.11$\% less than the one of the \Controller ($\mu_{controller} = 6.1mm$ \vs $\mu_{pen} = 5.3mm$).
A similar behavior was observed for \Hand input, where \Point gestures are $16.85$\% more precise compared to \Pinch gestures ($\mu_{pinch} = 8.9mm$ \vs $\mu_{point} = 7.4mm$).

\prepara
\noindent $\bullet$ \emph{\mbox{\textbf{[MN-2]}} {\bf NLM --}}~
Using a \Tool generates a significant higher number of NLM ($ 385.08$\%), compared to using \Hand gestures ($\mu_{tool} = 143.1$ \vs~ $\mu_{hand} = 29.5$, see Fig.~\ref{fig::eval_main_effect}b). 
Recall that \mbox{[MN-1]} implied that when using a \Tool, users generate more accurate drawings than using \Hand, which is possibly caused by more frequent attempts to correct errors.
To verify this, we normalized the time series instantaneous stroke--target distance~(\aka~error) using $z$--score normalization, followed by a Fast Fourier Transform~(FFT). 
Magnitudes of each frequency bins~(with $0.1$~Hz resolution) were aggregated by averaging the values across all tasks and participants with the same input methods.
As evident in Fig.~\ref{fig::eval_main_effect}j, the FFT of instantaneous error generated by \Tool is significantly higher than that yielded by \Hand, potentially causing the traces to look more fluctuating.
While Fig.~\ref{fig::eval_main_effect}b shows a slightly higher NLM when comparing \Pen \vs~\Controller, and \Point \vs~\Pinch, the pairwise difference between \Pen and \Controller does not show statistical significance. 

\prepara\noindent $\bullet$ \emph{\textbf{[MN-3]} {\bf MSP --}} When looking at MSP, drawing with \Controller ($\mu_{controller} = 11.89cm/s$) and \Pinch ($\mu_{pinch} = 12.54cm/s$) is significantly faster than that using \Point ($\mu_{point} = 10.84cm/s$), followed by \Pen ($\mu_{pen} = 8.48cm/s$) (see Fig.~\ref{fig::eval_main_effect}c). 
%
%
One possible reason is the different use of finger dexterity across different devices. 
With \Pen and \Point, participants were encouraged to use thumb and index finger, as well as index finger only, respectively, to control and stabilize the strokes.
While these could enhance the precision, the average drawing speed would be sacrificed.
In contrast, with \Controller and \Pinch, participants were encouraged to use the lower arms to control the stroke movement, which could benefit drawing speed, yet sacrifice accuracy.

\prepara
\noindent{\bf Curvatures \& Slope --}
RM-ANOVA shows that curvature has a significant impact on MOD ($F_{2, 22} = 88.80, p < .001, \eta^2_p = 0.131$), NLM ($F_{2, 22} = 47.21, p < .001, \eta^2_p = 0.074$), and MSP ($F_{2, 22} = 79.93, p < .001, \eta^2_p = 0.120$).
We also show that the slope measure only statistically affects NLM ($F_{2, 22} = 6.80$, $p < .01$, $\eta^2_p = 0.011$). 
Fig.~\ref{fig::eval_main_effect}~(d -- g) illustrate key takeaways:


\prepara
\noindent$\bullet$\emph{\mbox{\textbf{[MN-4]}} {\bf MOD \& NLM \& MSP -- }} An increase in curvature generally reduces the MOD ($\mu$, $8.0mm$, $6.6mm$ and $6.0mm$ for \SMA, \MED and \LAG, respectively).
While examining Fig.~\ref{fig::eval_main_effect}e, a lower NLM is visible when the curvature is larger ($\mu$, $90.79$, $85.92$, and $82.25$ for \SMA, \MED and \LAG respectively, see Fig.~\ref{fig::eval_main_effect}e).
This possibly means that participants tend to draw slower with increasing curvature, as evident in Fig.~\ref{fig::eval_main_effect}f ($\mu$, $11.99cm/s$, $10.92cm/s$, and $ 9.92cm/s$ for \SMA, \MED and \LAG, respectively).

\prepara
\noindent$\bullet$ \emph{\mbox{\textbf{[MN-5]}} {\bf NLM --}} While Fig.~\ref{fig::eval_main_effect}g shows that an increase in slope reduces NLM, only pairwise comparisons of \SMA and \LAG trials show statistical significance ($p < .01$), and therefore the contributions from the Slope are overall quite weak.

%

\prepara
\noindent{\bf Orientation --}
RM--ANOVA shows that orientation has statistically significant impacts on MOD ($F_{2, 22} = 24.16$, $p < .001$, $\eta^2_p = 0.039$) and NLM ($F_{2, 22} = 26.77$, $p < .001$, $\eta^2_p = 0.044$).
%

\prepara
\noindent$\bullet$\emph{\mbox{\textbf{[MN-6]}} {\bf MOD \& NLM -- }} Participants generated lower MOD and higher NLM for \FB~and \LR orientations, compared to the \UD trial ($\mu_{MOD}$ = $6.50mm$, $6.55mm$, $7.65mm$ and $\mu_{NLM}$ = $88.73$, $88.65$, $81.42$ for \FB, \LR, and \UD, post-hoc contrasts, with $p<.001$, see Fig.~\ref{fig::eval_main_effect} (h -- i)).
This implies that orienting the helix in \FB and \LR configurations helps enhancing 3D drawing precision.
%



\presubsubsec
\noindent
\underline{\bf Analysis of Interaction Effects}~\\
Interaction effects refer to those when one factor on the measure depends on the level of the other factor.
Fig.~\ref{fig::interaction_effects} summarizes the two--way interaction effects that are statistically significant~($p < .05$).
For the post-hoc contrast test, we used, \eg~ $A_iB_j \sim A_kB_l$, to represent the pairwise comparison between $A_iB_j$ and $A_kB_l$, where $A_i$, $A_k$ and $B_j$, $B_l$ are possible levels of factor $A$ and $B$. 
When $j = l$, we use the notation of $A_i \sim A_k \mid B_j$.

\prepara
\noindent{\bf Input Method + Curvature --}
Interaction effects between curvature and input methods on MOD ($F_{6, 66} = 2.70$, $p < .05$, $\eta^2_p = 0.014$) and NLM ($F_{6, 66} = 27.34$, $p < .001$, $\eta^2_p = 0.122$) are identified in Fig. ~\ref{fig::interaction_effects}(a -- b), and summarized below:

%
\prepara\noindent $\bullet$
\emph{\mbox{\textbf{[ITA-1]}} {\bf MOD \& NLM for Tools -- }}~
\Pen generates more precise drawings than \Controller, but only at at larger curvature tasks (\Pen $\sim$ \Controller at \SMA, \MED, and \LAG conditions: $p >.05$, $p < .01$, and $p <.001$, see Fig.~\ref{fig::interaction_effects}a).
While interactions with the \Pen also introduce higher NLM compared to using the \Controller at larger curvature, the post-hoc contrast tests do not show statistical significance (see Fig.~\ref{fig::interaction_effects}b).
%
This verifies our finding in Fig.~\ref{fig::eval_main_effect}b. 

\prepara\noindent $\bullet$
\emph{\mbox{\textbf{[ITA-2]}} {\bf MOD \& NLM for Hand -- }}~
When using a \Hand gesture while drawing, we found that the \Point gesture generates significantly less error compared to the \Pinch one, but only in \SMA curvature tasks ($p < .05$), while no statistical significance is detected during \MED and \LAG curvature trials.
Additionally, the \Point gesture generates significantly higher NLM in \SMA and \MED curvature tasks compared to the \LAG trials (\Pinch $\sim$ \Point in \SMA, \MED, and \LAG curvatures: $p <.001$, $p<.05$, $p >.05$).

\prepara
\noindent{\bf Input Method + Orientation --}
RM--ANOVA results in Fig. ~\ref{fig::interaction_effects}~(c -- d) show the interaction effects of input methods and orientations on MOD ($F_{6, 66} = 4.47$, $p < .05$, $\eta^2_p = 0.022$) and NLM ($F_{6, 66} = 4.80$, $p < .05$, $\eta^2_p = 0.024$). In short:

\prepara\noindent $\bullet$
\emph{\mbox{\textbf{[ITA-3]}} {\bf MOD \& NLM for Tools -- }}
Drawing with a \Tool in general, and the \Pen specifically produced significantly higher precision during \textit{Left--Right} orientation ($p < .01$). The other two orientations, did not show any advantages when using the \Pen.

\prepara\noindent $\bullet$
\emph{\mbox{\textbf{[ITA-4]}} {\bf MOD \& NLM for Hands -- }}~
When using a \Hand gesture, we found that participants had a tendency of being more precise while drawing using \Point than \Pinch, but only in the \textit{Left--Right} orientation. The difference between the two \Hand modalities however was not statistical significant ($.05 < p < .1$).

\prepara
\noindent{\bf Orientation + Curvature / Slope -- }
Fig.~\ref{fig::interaction_effects}{e} shows the RM--ANOVA results for the interaction effects of curvature and orientation on MOD ($F_{4, 44} = 4.23$, $p < .01$, $\eta^2_p = 0.014$). 
And Fig. ~\ref{fig::interaction_effects}~(f--g) show the interaction effects of orientation and slope on MOD ($F_{4, 44} = 3.54$, $p < .01$, $\eta^2_p = 0.012$) and NLM ($F_{4, 44} = 6.08$, $p < .05$, $\eta^2_p = 0.020$).

\noindent$\bullet$\emph{\mbox{\textbf{[ITA-5]}} {\bf MOD \& NLM -- }}~In the \SMA curvature condition, participants generated more precise output when drawing in the \FB and \LR orientation.
As for slope, the \UD orientation produces the least accurate trials in all slope conditions, and the \SMA slope generates significantly lower NLM in the \UD~($p<.001$) orientation, compared to the other two orientations (see Fig.~\ref{fig::interaction_effects}f -- g)


\prepara
\noindent{\bf Curvature + Slope -- }
Finally, Fig.~\ref{fig::interaction_effects}~(h--i) show the interaction of curvature and slope on MOD ($F_{4, 44} = 3.31$, $p < .05$, $\eta^2_p = 0.011$) and NLM ($F_{4, 44} = 5.56$, $p < .001$, $\eta^2_p = 0.019$).

\noindent$\bullet$
\emph{\mbox{\textbf{[ITA-6]}}~{\bf NLM -- }}
%
%
The \LAG curvature and \LAG slope yields lowest NLM compared to the rest of cases.


\subsection{Uncontrolled and Intermediate Variables}\postsec\label{sec::results::intermediate}
It is also worth to investigate the correlation with critical intermediate variables that have been examined in prior works~\cite{FittsLaw, Isokoski2001, Accot1997}.
We base our description of the impact of intermediate variables on a causal model (see Fig.~\ref{fig::eval_intermediate}a) with three hypothesized Causal Links~(CLs), considering two specific intermediate variables:

\prepara
\noindent{\bf ($1$) \underline{S}troke--\underline{S}houlder \underline{D}istance (SSD) --}
%
SSD is defined as the Euclidean distance between shoulder and real-time stroke position.
We assumed that SSD can affect drawing performance (see Fig.~\ref{fig::eval_intermediate}a). 
%
Since the arm length varies among participants, we considered the percentage of absolute SSD over participants' arm length.

\prepara
\noindent{\bf ($2$) \underline{A}bsolute Drawing \underline{SP}eed (ASP) -- }
While we instructed participants to prioritize precision over speed, actual speeds could vary within each trials and across the participants depending on multiple confounding factors, such as the use of specific tools, and specific task geometric features~\cite{Arora2017}.

Fig.~\ref{fig::eval_intermediate}~(b--d) show the correlations among SSD, ASP and stroke-target distance (\ie~error).
We used Spearman's Rank-Order Correlation to examine the monotonic relationship between two variables. We outline the key findings below:

\prepara
\noindent\emph{\textbf{[ITM--1, CL1] ASP \vs~SSD --}} 
We found only a very weak monotonic correlation between distance and speed for \Hand gestures ($p > .1$).
When using \Hand gestures, participants draw faster when the strokes are either far or very close to their shoulder.
In contrast, when using a \Tool, the drawing speed increases with higher values of SSD for both \Pen and \Controller ($r_s(7) = 0.95$, $p < .005$).

\begin{figure*}[t]
    \centering
    \includegraphics[width=\textwidth]{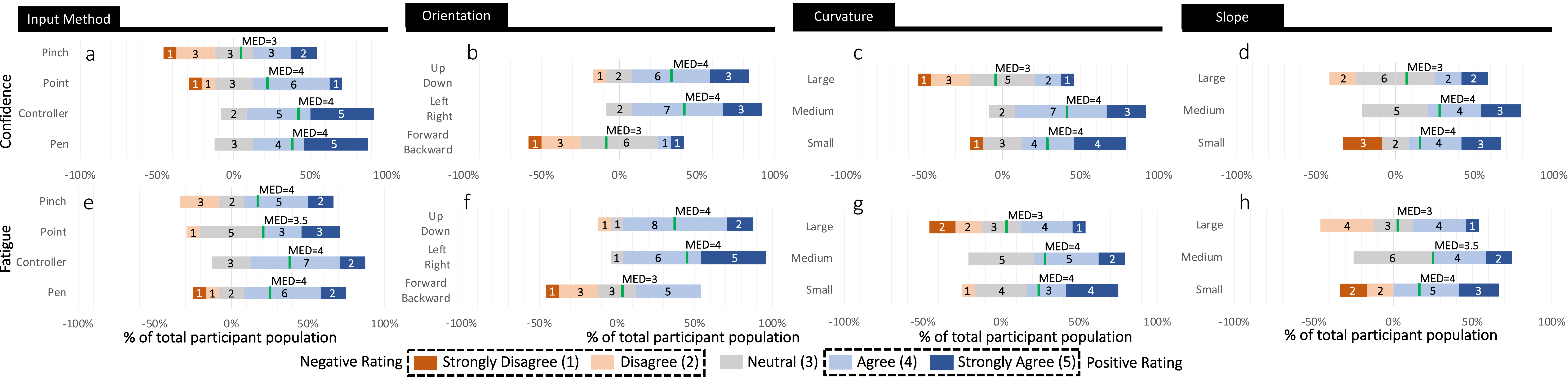}
    \vspace{-0.30in}
    \caption{Participants' answers to post-study questionnaires. The Likert scores corresponding to the level of agreement are $1$ -- $5$. We use the green line to mark the mid point with the label of median~(MED).}
    \vspace{-0.23in}
    \label{fig::ratings}
\end{figure*}

\prepara
\noindent\emph{\textbf{[ITM--2, CL2] Error \vs~SSD -- }}
When looking at error rates and their relationship with distance, we observed a strong monotonic correlation for both \Pen ($r_s(7) = 0.93$, $p < .005$) and \Controller ($r_s(7) = 0.82$, $p < .01$), while such correlation was very weak for \Hand ($p > .1$). 
When using \Hand gestures, with SSD $\sim 60\%$ of the length of their arm, the error is minimized.
When interacting with a \Tool, with SSD $30\% \sim 40\%$ of the length of their arm, error is minimized.

\prepara
\noindent\emph{\textbf{[ITM -- 3, CL3] Error \vs~ASP --}}~
We observed a strong monotonic correlation between speed and error for all input modalities (\Controller: $r_s(14) = 0.84$, \Pen: $r_s(14) = 0.96$, \Pinch: $r_s(14) = 0.79$, \Point: $r_s(14) = 0.73$; $p < .005$ for all modalities). 
Fig.~\ref{fig::eval_intermediate}c also shows that the optimal speed for achieving minimal stroke--target distance for \Controller, \Pen, \Pinch, and \Point is around $6cm/s$, $2cm/s$, $5cm/s$ and $2cm/s$ respectively.

\subsection{Post--Study UX Evaluations}\postsec\label{sec::results::subjective}
We analyzed the Likert--scale results of the questionnaire, and participants' comments.
%
%
We used median~(MED) for analysis due to the ordinal nature of Liker--scale data~\cite{Sullivan2013, Jamieson2004} (Fig.~\ref{fig::ratings}).
%
%
Thematic analysis was used to identify key takeaways from qualitative responses.
%

%
\prepara
\noindent\textbf{Using \Tool yields high--confidence with the drawings, however using the \Pen might introduce higher fatigue than using the \Controller ~\textbf{\mbox{[SUB-1]}}}

\noindent Most participants felt more confidence while using a \Tool compared to using \Hand gestures (see Fig.~\ref{fig::ratings}a).
%
%
Two participants pointed out how they were more confident to use the \Controller for balancing the speed--accuracy trade-off, while they preferred the \Pen for accuracy only:
{\it``For the controller, [...] I can go faster while maintaining reasonable accuracy. But for the pen [...] it has a tip, so I can mentally project the lines, even if the tip is small.''}\enspace [P8].

Two participants reported that using the \Pen would result in increased fatigue, while no participants specifically pointed to it for the \Controller.
%
%
Possible reasons includes the \Pen's heavy weight (\eg~{\it``It's a little bit heavy in the back. I felt that I struggled to hold it up-right.''}\enspace [P2]), its bulky size (\eg~{\it ``[...] so thick that I cannot have a good grip.''}\enspace [P9]),  its non-ergonomic gripping gestures (\eg~{\it``[...] the positioning is a little bit odd...like if this was a normal pen, I would have hold it closer to the back.''}\enspace [P2]
), 
and its instability, caused by the stroke triggering actions (\eg~{\it``[...] pressing down [the drawing button] influenced my accuracy [...]. I assume it is because I am tapping against the pivot point.''}\enspace [P8]).

\prepara
\noindent{\textbf{Using \Hand Input decreases fatigue, but the reason varies among \Pinch and \Point \textbf{[SUB-2]} --}}
Participants reported how they felt that {\it``the hand type of gestures would have the lowest fatigue [compared to the tools], because [of] not holding anything''}\enspace [P2].
%
%
While the median fatigue rating of \Pinch is slightly higher than \Point ($4$ \vs$3.5$), three participants gave a \textit{Negative} rating for \Pinch, while only one rated \Point negatively.
Possible reasons for feeling increased fatigue while using both hand gestures include the needs of mental attention for \Pinch~(
{\it ``[the pinching gesture] requires me to mentally force to pinch, so this might cause slightly higher fatigue than pointing.''}\enspace [P2]) 
and the additional stretching of fingers for the \Point gesture ({\it ``The reason of increased fatigue with hand pointing is that I have to keep my finger stretched.''}\enspace [P10]).

\prepara
\noindent{\bf Occlusions and poor hand tracking hinder confidence \textbf{\mbox{[SUB-3]}}}~\\
Confidence greatly decreased when participants were asked to use \FB orientation, \LAG or \SMA curvature (see Fig.~\ref{fig::ratings}b, c).
%
%
Most likely, the awkward positioning of the geometric shapes of these tasks, caused participants to not see the target traces in their entirety (\eg~{\it``sketching the smallest slope cases, it was really hard to distinguish the traces to make the next move.''}\enspace[P8]). 
%
%
%
%

%
Four and two participants gave \textit{Negative} confidence rating for the \Pinch and \Point respectively, but not for any of the two \Tool~(Fig.~\ref{fig::ratings}a).
Participants explained this mostly in terms of the \Hand tracking performance, that felt less precise as compared to the \Tool, causing less confidence while controlling strokes for precise drawing (\eg~{\it``I think hand tracking is generally a big problem since [sometimes] the hand tracking is going crazy. Whereas for the controller and pen, they were always tracked perfectly.''}\enspace [P8], {\it ``the tracking failures causing the disappearing of [the] virtual hands [hindered] keep going with the previous gesture.''}\enspace [P10]). 
%
%

%
\prepara
\noindent\textbf{\MED curvature \& slope yield best delineation experiences \textbf{[SUB-4]} --}
Despite \LAG curvature generates higher precision (see \mbox{[MN-4]}), we observed that the \MED curvature and slope typically obtain the highest number of \textit{Positive} and lowest number of \textit{Negative} qualitative ratings (Fig.~\ref{fig::ratings}c, d).
Participants reported how \LAG slope and curvature require more attention and fine grained control, which might cause fatigue and reductions of both drawing speed and precision: {\it ``When the slope is large, I am deviating a lot and need more effort and attention to control the stroke''}\enspace [P8],
{\it ``when [curvature] is large, you cannot really make your movement fast enough.''}\enspace [P2], and
{\it ``[large curvature] was harder to get it more precise.''}\enspace [P11].
On the other hand, \SMA curvature could lead to fatigue: {\it ``requires a lot of your arm movement, and it could be more tiring.''}\enspace [P2]), and could cause more occlusion [P8].

\prepara
\noindent{\textbf{Interactions between curvature and orientation could affect the delineation experience \textbf{[SUB-5]}--}}
While we only consider the main effects, we found that some participants felt that the interactions of curvature and orientations also had important effects on the drawing experience.
For instance, P8 commented: 
{\it``for the forward-backward configuration, when [the tasks] are large [\ie~small curvature], it was super easy. When they are small [\ie~large curvature], I am not so sure.''}, and: {\it`for the left-right configurations, when they are large [\ie~small curvature], it is super hard, but when they are small [\ie~large curvature], it was pretty easy''}.

%% file: 05-discussion.tex
\section{Discussion and Design Recommendations}\postsec\label{sec::discussions}
%
%
%
%

\subsection{Pre--Session Design Recommendations}\postsec\label{sec::discussion::presession}
Pre--session indicates the strategies that are considered {\it before} each drawing session, which could be summarized into two aspects:

\prepara
\noindent{\bf VR Stylus is recommended for precision--first 3D drawing.}
%
\noindent We showed that using \Tool, in particular \Pen, yields better overall drawing precision compared to other approaches with highest frequent attempts for error correction.
%
%
Although bare hand drawing means not having to hold any physical objects, participants felt less confident when performing precise--first drawings. primary because of poor hand tracking performance (see \mbox{[SUB-3]}).
%
%
%
%
%
Unlike \Tool--based approaches, whose positional tracking is achieved by measuring the time--of--flight of IR pulses emitted by lighthouses, hand tracking is realized using machine learning solutions~\cite{OculusHandTrackingPrinciple, Han2020}.
This resulted in a few inherent issues where certain hand poses could not be well tracked, yet users needed them to perform precision--first 3D drawing. This is most likely due to limited sample diversity of the original training dataset. 
%
%
Although using bare hand is widely advertised by vendors~(\eg~\cite{OculusHandTrackingGestures}) as a better modality for 3D interactions (see Sec.~\ref{sec::related::tools}), we showed that using \Hand gestures to perform precise 3D drawings in today's VR headsets is not as accurate as using a \Controller.
Therefore, designers can benefit from utilizing a VR \Pen for precise 3D drawing, and allocating coarse--grained tasks (\eg~target selections) to \Hand.

\prepara
\noindent{\bf Addressing occlusion could result in increased drawing precision.}
%
%
The source of occlusions includes the occlusions by the drawing tasks itself and the photo--realistic prefab of drawing tools.
%
%
%
%
While entirely removing the photo-realistic hand or tool visual images in VE might solve this issue, it might reduce users' proprioception, and ultimately could reduce the overall sense of immersion~\cite{Gallagher2000, Gonzalez2018}.  
%
%
%
Instead, we believe that new interaction techniques can extend current work that focus on, for example, moving the physical hand position away from the stroke position (\eg~\cite{Pfeuffer2015}) or optimize the transparency of photo-realistic prefab similar to~\cite{Auda2021}.

%

\subsection{In-Session Design Recommendations}\postsec\label{sec::discussion::insession}
%
%
In-session recommendations refer to the design opportunities {\it during} the drawing session. 
While geometry is usually unknown prior to the practical drawing experience and requires designers to continuously adapt to the stroke characteristic---which in turns breaks their workflow---we believe that our in-session design recommendations promote feedback that could potentially intervene in real-time and help participants decide on {\it whether} and {\it how} to adjust their behaviors and transform the drawing canvas to obtain better results. 
%

\prepara
\noindent{\bf Fatigue, drawing speed, stroke--shoulder distance, and hand tracking precision should be assessed in real-time.}
%
While using the \Pen yields higher precision overall, this does not mean that the \Pen can be blindly used in all drawing scenarios.
%
Ergonomically, we indicated how using the \Pen might introduce higher fatigue compared to using a \Controller~\mbox{[SUB-1]}.
%
%
While a \Pen could support more precise 3D drawing, the use of thumb, index and middle fingers seemed to reduce the overall drawing speed and induce more fatigue.
%
In contrast, the use of a \Controller seems to be more beneficial for fast and large movement (\eg~while drawing the \SMA curvature), compared to fine-grained movements (\eg~while drawing the \LAG curvature). 
Although {\it``control and precision''}\enspace have been advertised as the highlights for some commercially available VR stylus~\cite{LogitechVRInk}, we implied that this might not apply for \SMA curvature and \FB or \UD orientations.
%
\Second, our results also showed the impacts from SSD and ASP~(see Sec.~\ref{sec::results::intermediate}).
Although participants were prompted to prioritize drawing precision, they tended to increase the drawing speed when they felt confident, they perceived the current drawing modality to be more convenient, and they experienced less fatigue (\mbox{[SUB-1]}). 
This, however, might led to less precise drawings, even though {\it ``precision--first''} was used as the prompt throughout the study.
%
%
Although parameters such as the geometry of the drawing tasks are hard to predict and control, being aware of and capable of regulating the stroke speed~(ASP) and stroke distance~(SSD) might be helpful to improve precision.
%
Therefore, we recommend to integrate real--time visual feedback to regulate participants ASP, SSD, and the time for resting to minimize the impacts of fatigue.
This is different from prior research that tried to use different forms of feedback (\eg~visual anchors~\cite{Arora2017, Machuca2019}) for assisting participants ``feel'' the spatial characteristic (\eg~the sense of depth).
\Finally, when it comes to hand, we showed that occasional poor hand tracking might be one main reason for causing low confidence [SUB-3]. 
While this is limited by today's tracking technology, we recommend to also integrate the visual feedback for real-time hand tracking precision.
Such visual prompts could encourage participants to {\it in--situ} adjust their gestures to maintain high drawing performance while conducting precision--first 3D drawing.

\prepara
\noindent{\bf Real--time visual feedback could assist to find optimal scale and orientation of the workspace.}
We indicated that the uses of \Pen and \Point gestures achieve more accurate drawings in \LR orientations.
%
This could be possibly due to participants' mental model of drawing on a 2D surface (similar to using pen-and-paper), in which a stylus is usually held perpendicular to the drawing surface~\cite{Fitzmaurice1999}.
%
%
%
%
%
However, when presented with orientations different from \LR, participants needed to find other ways to perform precise drawing: they changed the directions of \Pen or \Point by rotating their wrists, which resulted in breaking the drawing workflow and might have caused  discomfort, reduced drawings' quality~\cite{Fitzmaurice1999}, and increased cognitive and sensorimotor load~\cite{Wiese2010}.
%
%
Additionally, while \LAG curvature resulted in higher precision, it demanded finer control and prevented participants to draw fast.
This possibly contributed to participants considering \MED curvature as the optimal option in terms of convenience and confidence in their subjective evaluations. 
While we only consider curvature and slope as categorical variables with three possibilities, these variables are continuous in practical 3D modelling systems.
%
While the geometry of the tasks is unknown prior to the drawing, future system could  analyze the directions and the velocity of the previously drawn strokes, and used them to ``guess'' the direction of future strokes.
A real--time visual feedback containing guidance for optimal workspace orientations and scale, could lead to better curvature and slope.

%% file: 06-limitation.tex
\section{Limitations}\postsec\label{sec::limitation}
We summarize our limitations into four main categories.

\noindent{\bf (1)}~Our participants showed above--average spatial abilities, which might affect some results on overall abilities to make precise delineations.
Future studies can consider more participants with a more diverse spatial abilities and VR experience.

\noindent{\bf (2)}~To minimize complexities, participants were only asked to report on main interaction effect qualitatively during post--study evaluation.
However, P8 found it difficult to provide rate on several questions, and mentioned the needs for more dimensions.
Future works might consider evaluating subjective experience for interaction effects, and potentially include a post--study workshop~\cite{Yarmand2021} that can facilitate sharing more thoughts and articulate feelings more in--depth.

\noindent{\bf (3)}~While we intentionally included breaks into task design to mitigate fatigue, and this has been acknowledged positively by some participants, other participants extensively referred to fatigue during post--study UX evaluations. It is not clear how much of the reported fatigue is due to the experiment design and how much to the tasks themselves.
Future studies might consider objective and quantifiable measures of the level of fatigue throughout the study~\cite{Shamli2021}. 

\noindent{\bf (4)}~Our study used Oculus Quest $2$~\cite{OculusQuest} due to its affordability and commercial availability, which leverage a deep learning based approach to track hands. 
Besides human factors, the less accurate hand tracking performance, compared to the tools, might also contribute to the low MOD~[MN-1]. 
While this is considered as part of our study, future work might also consider the individual impacts of tracking accuracies caused by technology imperfections. 

\noindent{\bf (5)}~We considered the geometric characteristics in our study as categorical variables with only three possibilities for each of the factors.
However, practical drawing characteristics are not categorical, but typically are parameterized by numerical variables. 
Although we believe that the three categories we chose are a good representation of the possible spectrum, future works might decide to investigate and establish more precise mathematical models between the specific numerical variables and the drawing performance outcomes.

%% file: 07-conclusions.tex
\section{Conclusion}\postsec\label{sec::conclusion}
We demonstrated how geometric characteristics and input methods can affect precision--first 3D drawing in VR.
Through a within--subject study \mbox{(n = \N)}, we found that using a VR \Pen and \Controller leads on average to $29.63$\% more precision compared to using \Hand, and that the \Pen yields $13.11$\% higher precision than the \Controller.
We showed how \LAG curvature, \FB or \LR orientations allow participants to generate more precise drawings.
Together with participants' user experience, we finally proposed four recommendations for enhancing accuracy and experience of precision--first 3D drawing that could be integrated in future 3D immersive drawing tools.

%% file: appendix.tex

\renewcommand{\thefigure}{A\arabic{figure}}
\setcounter{figure}{0}
\section{Additional Modeling and Data}
\subsection{Modelling of the Task Traces}\label{sec::app::helix}
Mathematically, a helical curve $\textbf{c}(t; r, a)$ can be described parametrically by equation~(\ref{eqn::helical_curve}), where curvature ($k$) is independent to $t$ and decided by radius ($r$) (see equation~\ref{eqn::helical_curvature}) and slope ($s$) (see equation~\ref{eqn::helical_slope}).

\begin{equation}
    {\bf c}(t; r, a) = [rcos(t), rsin(t), at]
    \label{eqn::helical_curve}
\end{equation}

\begin{equation}
    k = \frac{\left\Vert{\bf \hat{c}}''(t)\right\Vert}{\left\Vert{\bf c}'(t)\right\Vert} = \frac{r}{r^2 + a^2}
    \label{eqn::helical_curvature}
\end{equation}

\begin{equation}
    s = \frac{a}{r}
    \label{eqn::helical_slope}
\end{equation}

\begin{figure}[ht]
    \centering
    \includegraphics[width=0.3\textwidth]{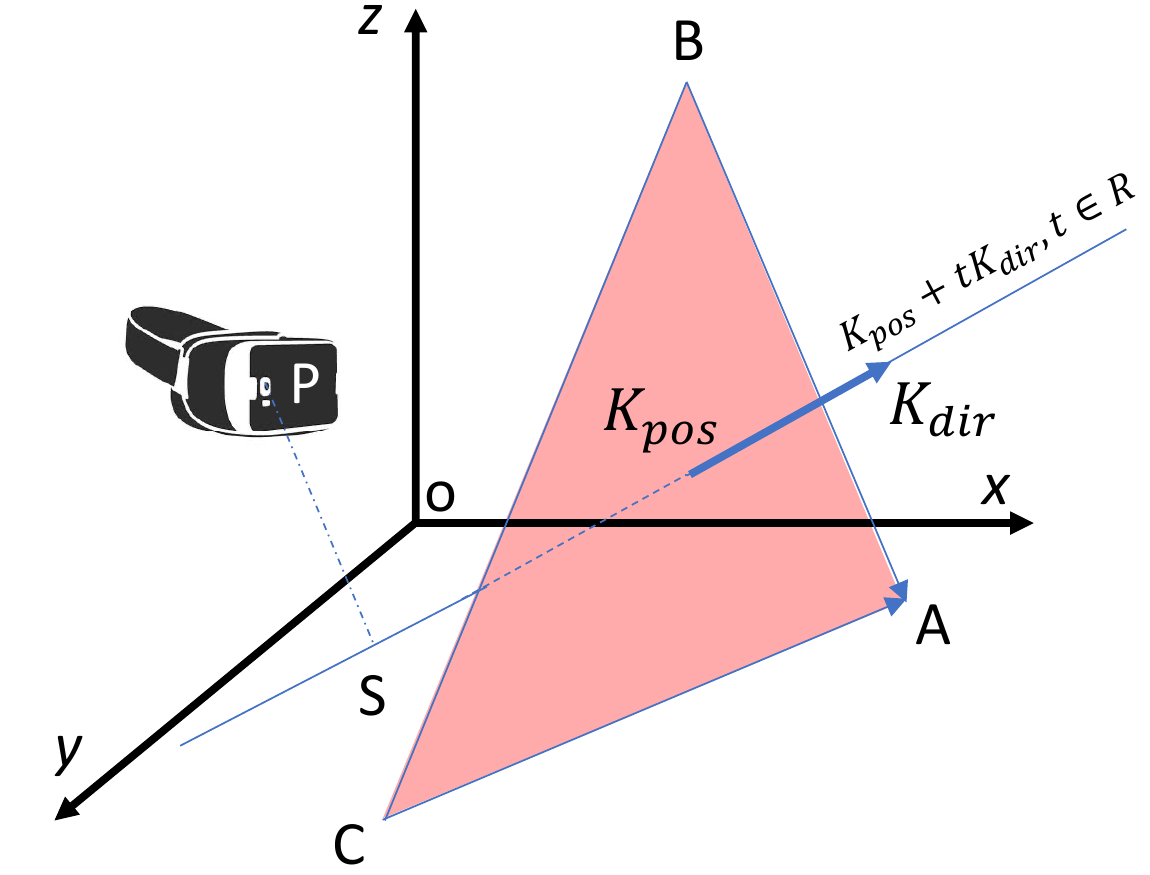}
    \vspace{-0.30cm}
    \caption{Illustrative diagram of mathematical modelling of the calibration results.}
    \label{fig:my_label}
\end{figure}

By transforming (\ref{eqn::helical_slope}), and putting into (\ref{eqn::helical_curvature}), we can generate (\ref{eqn::helical_slope_k_to_r}) and (\ref{eqn::helical_slope_final_r}).

\begin{equation}
    k = \frac{r}{r^2 + s^2r^2} = \frac{1}{(1 + s^2)r}
    \label{eqn::helical_slope_k_to_r}
\end{equation}

\begin{equation}
    r = \frac{1}{(1 + s^2)k}
    \label{eqn::helical_slope_final_r}
\end{equation}

Finally, we can get (\ref{eqn::helical_slope_final_a}) by considering (\ref{eqn::helical_slope_final_r}) and (\ref{eqn::helical_slope}).

\begin{equation}
    a = sr = \frac{s}{(1 + s^2)k}
    \label{eqn::helical_slope_final_a}
\end{equation}

Plugging (\ref{eqn::helical_slope_final_r}) and (\ref{eqn::helical_slope_final_a}) into (\ref{eqn::helical_curve}), we are able derive the mathematical modelling, depending on $k$ and $s$ for generating task with predefined parameters.

\begin{equation}
    {\bf c}(t; k, s) = [\frac{cos(t)}{(1 + s^2)k}, \frac{sin(t)}{(1 + s^2)k}, \frac{st}{(1 + s^2)k}]
    \label{eqn::helical_curve_final}
\end{equation}

After this, the generated sample points would then be transformed (\ie~shifted and rotated) based on the calibration results at the beginning of each study session.
The calibration process would compute and approximate the shoulder positions from $3$ controller locations while being placed in the position of forward, up and right.
Assuming the sample points at the positions of front, up and right are $A$, $B$, and $C$, we can compute the position of the circumcenter ($K_{pos}$) and normal direction ($K_{dir}$) of the triangle $\Delta ABC$ (see Fig.~\ref{eqn::shoulder_pos})~\cite{find_circumcenter}.

\begin{equation}
\begin{split}
K_{pos} = A 
    &+ \frac{||C - A||^2[(B - A)\times(C - A)] \times (B - A)}{2||(B -A) \times (C - A)||^2} \\
    &+ \frac{||B - A||^2[(C - A)\times(B - A)] \times (C - A)}{2||(B -A) \times (C - A)||^2}
\end{split}
\label{eqn::circumcenter_pos}
\end{equation}

\begin{equation}
    K_{dir} = \overrightarrow{BA} \times \overrightarrow{CA}
    ~\label{eqn::circumcenter_dir}
\end{equation}

Assuming $P$ is the position of the headset in world coordinate, we can then approximate the shoulder position ($S$) by finding the closest point on line $K_{pos} + t K_{dir}$ ($t \in \mathbb{R}$) (see (\ref{eqn::shoulder_pos})).

\begin{equation}
    S = K_{pos} + \frac{K_{pos} + K_{dir}}{||K_{pos} + K_{dir}||} \frac{\overrightarrow{K_{pos}P} \cdot (K_{pos} + K_{dir})}{||K_{pos} + K_{dir}||}
    \label{eqn::shoulder_pos}
\end{equation}

\subsection{RM--ANOVA Significance Test Results for Main and Interaction Effects}\label{sec::anova_results}
\begin{figure*}[h]
    \centering
    \includegraphics[width=0.8\textwidth]{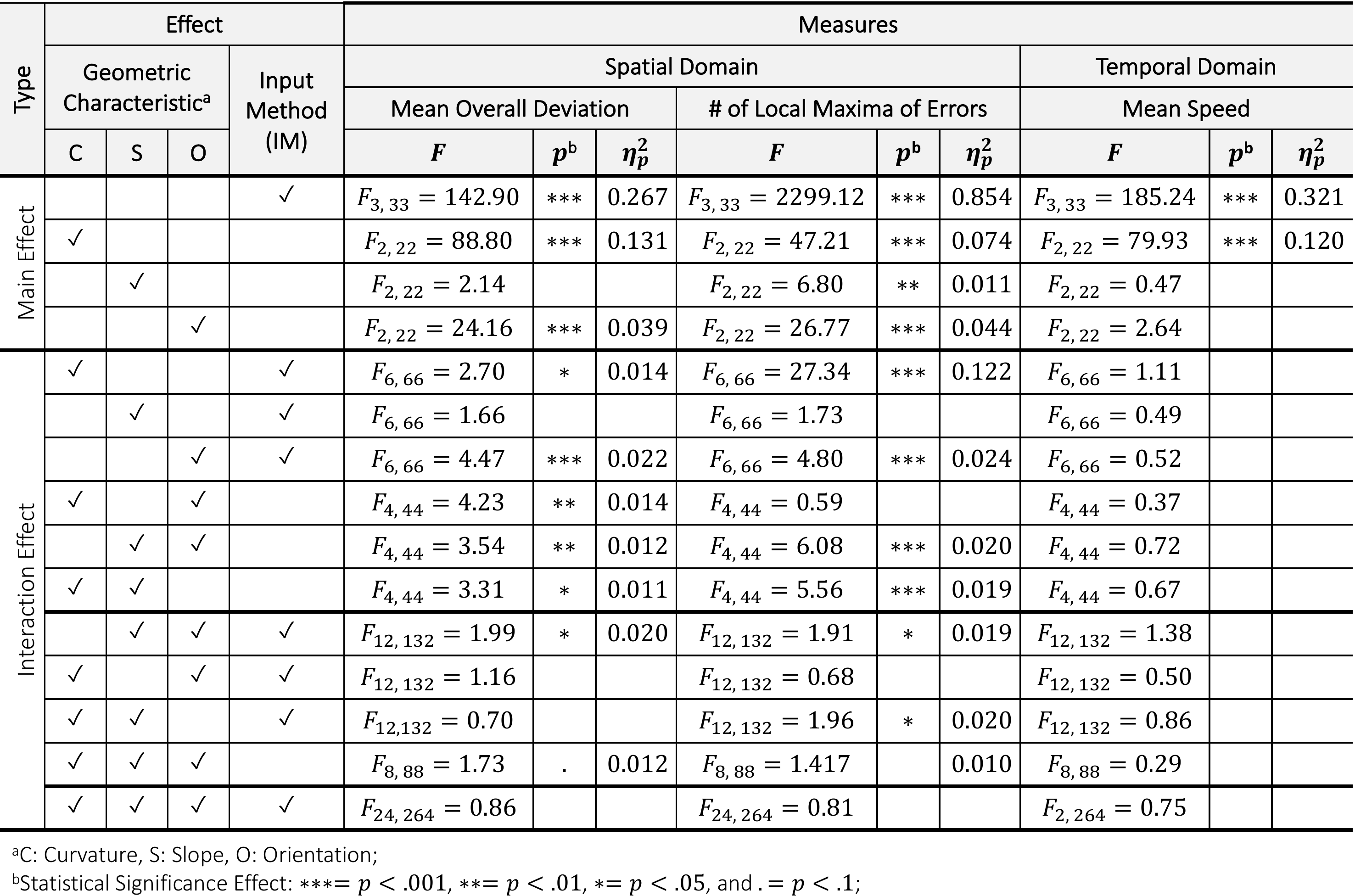}
    \vspace{-0.05in}
    \caption{Results of repeated measure ANOVA.}
    \label{fig::anova_results}
\end{figure*}

The full results of RM-ANOVA computations can be seen in Fig.~\ref{fig::anova_results}. %
In \S\ref{sec::evaluation}, the post-hoc contrast tests were conducted \textit{only if} the main or interaction effects are statistical significance ($p < .05$).

\subsection{Results without aggregated analysis}\label{sec::app::full_results}
\begin{figure*}[h!]
    \centering
    \includegraphics[width=\textwidth]{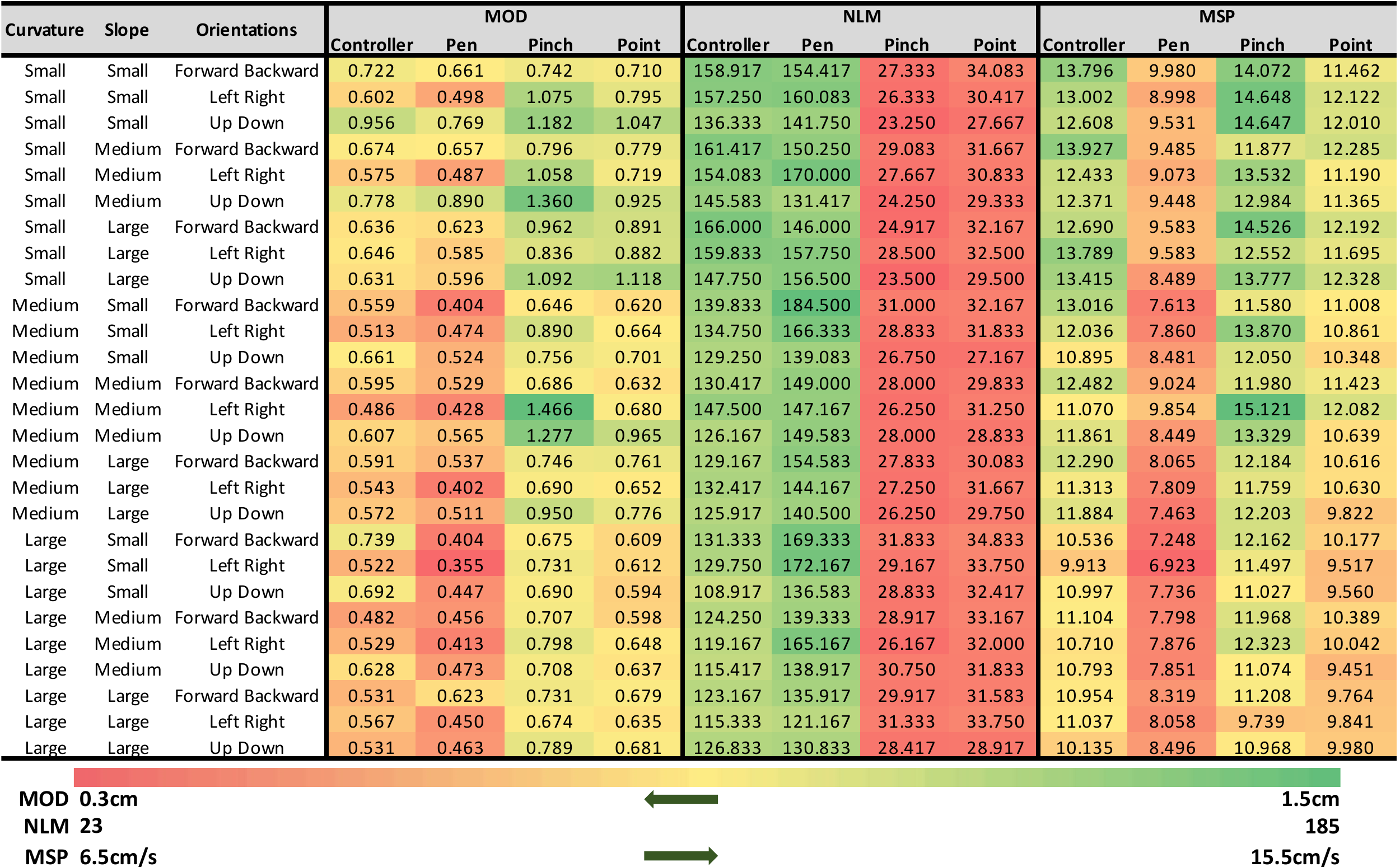}
    \vspace{-0.2in}
    \caption{MOD, NLM and MSP of each of $108$ trials, averaged across \N~participants. The green left\/right arrows at bottom legend indicate the target performance preference.}
    \label{fig:all_tasks_results}
\end{figure*}

In \S\ref{sec::results::independent}, we analyze the main and interaction effects of $4$ within--subject factors. 
In this note, we listed out the $3$ measures of each of $108$ trials, by averaging the results collected on each of \N~participants.  
The results are listed in Fig.~\ref{fig:all_tasks_results}.

\subsection{Dataset}\label{sec::app::dataset}
To encourage future research, we have made our dataset publicly available\footnote{An interactive web--based visualizer of our dataset can be accessed here: \url{https://precision-first-vr.github.io}.}.
Although the focus of this work lies on the analysis of spatial and temporal features of the drawn traces, we also record participants participant's drawing behaviors. 
%
This unlocks future investigations for analyzing precision--first 3D drawing from the perspective of participants' drawing behaviors. 
Example key research questions include: 
{\it how do participants explore and observe during precision--first 3D drawing?}
and {\it how do participants hold the tools?}
\etc
This is one--step beyond the current method of analysis that only focuses on the temporal and spatial features of what participant drew.
To realize this, besides timestamp and sample world coordinates of participants' drawn strokes, we also recorded the transform\footnote{A transform in the world space is represented by the position vector and three Euler angles indicating rotation.} of tools (\ie~VR \Pen and \Controller) or hand, as well as camera in the world space, which indicates how participants looked at target of interest.